\newdimen\figwidth \figwidth= 7 truecm

\documentstyle[prl,aps,epsf,twocolumn]{revtex}

\newcommand{\vek}[3]{\left[\begin{array}{r}#1\\#2\\#3\end{array}\right]}

\begin{document}
\draft 
\title{Singularity confinement and chaos in discrete systems} 
\author{Jarmo Hietarinta${}^1$ and Claude Viallet${}^2$}
\address{${}^1$Department of Physics, University of Turku, FIN-20014
Turku, Finland\\ ${}^2$CNRS and Universit\'e Paris VI, Bo\^\i te 126,
4 place Jussieu, F--75252 Paris Cedex 05, France}
\maketitle
\begin{abstract}
We present a number of second order maps, which pass the singularity
confinement test commonly used to identify integrable discrete
systems, but which nevertheless are non-integrable. As a more
sensitive integrability test, we propose the analysis of the
complexity (``algebraic entropy'') of the map using the growth of the
degree of its iterates: integrability is associated with polynomial
growth while the generic growth is exponential for chaotic systems.
\end{abstract}


Discrete systems have for a long time been a subject of study in the
field of dynamical systems. A strong practical motivation is the power
of numerical exploration for such systems~\cite{henon}, which led to
interesting findings in chaos theory. On the other hand, several
numerical algorithms (``convergence acceleration algorithms'') are
related to integrable discrete maps (see \cite{nasu} and references
therein).  Finding out whether a given system is chaotic or integrable
is then a basic question worth more investigation.  One open problem
to find an {\em algorithmic} test of integrability, and this is the
subject of the present letter.

In the case of continuous systems one can test for the ``Painlev\'e
property''~\cite{kova,painleve,contp,conte}, which is closely related
to integrability and has considerable predictive power.  An analogue
of the Painlev\'e test for discrete systems was proposed
in~\cite{sing} and has been used as a powerful constructive tool,
e.g., to identify discrete Painlev\'e equations \cite{diskrp}.  This
{\em singularity confinement test} is similar in spirit to the
continuous Painlev\'e test in that it analyses behavior around a
movable singularity of the map. When a map is iterated, it may happen
that we reach a point for which the next value is ill-defined due to
the appearance of an indeterminate form, $\infty-\infty$,
$0\cdot\infty$ or such. One should then study the behavior around the
singularity: if the map can be continued in a way which allows, after
a finite number of steps, to exit from the singularity without loss of
information, then the system is said to pass the test.

We show in this letter that the confinement test is not sufficient to
ensure integrability. We also propose another indicator for rational
maps: a measure of the ``algebraic entropy''~\cite{BeVi97,FaVi93},
which has to do with global properties of the system (see later).

We shall consider the map
\begin{equation}
x_{n+1}+x_{n-1}=x_n+a/x_n^2,
\label{map}
\end{equation}
and some of its generalizations.  Relation (\ref{map}) defines a map
$(x_{n-1},x_n)\rightarrow x_{n+1}$.  The potential singularity of this
map is reached if, at some step (say step 0), we arrive at $x_0=0$
(with a finite nonzero $x_{-1}$), because then $x_1=\infty$,
$x_2=\infty$, but $x_3=\infty-\infty$ and it is not clear how to
proceed. To refine the analysis, let us assume that we arrive at
$x_0=\epsilon$ by $x_{-1}=u$, with suitable previous $x_n$'s.  With
these initial values we get the sequence (here $a$=1)
\begin{eqnarray*}
& x_{-1} &=  u , \\
& x_0 &= \epsilon,\\
& x_1 &= \epsilon^{-2}-u+\epsilon,\\
& x_2 &= \epsilon^{-2}-u+\epsilon^4+\dots,\\
& x_3 &= -\epsilon + 2 \epsilon^4+\dots,\\
& x_4 &= u-\epsilon+\dots.
\end{eqnarray*}
In this case the outcome is: ``$\infty-\infty=0$'' and the sequence
emerges from the singularity with the value $u$, i.e., without losing
the initial information.  This means that the system (\ref{map})
passes the singularity confinement test without problems; the
singularity structure $u\to \epsilon \to \epsilon^{-2} \to
\epsilon^{-2}\to -\epsilon\to u$ is rather typical.

The problem is that  system (\ref{map}) is chaotic, as we shall show. 
Our suspicion about the non-integrability of (\ref{map}) arose when,
motivated by~\cite{FaVi93}, we evaluated the growth of the degree of
its iterates as follows:

We start by writing the map as a first order two-dimensional map
\begin{equation}
\varphi: p_n=(x_{n-1}, x_{n}) \longrightarrow  p_{n+1}=(x_{n}, x_{n+1}),
\end{equation}
and then rewrite $\varphi$ in terms of homogeneous coordinates
$[y_n,z_n,t_n]$ by setting
\begin{equation}
p_n=\left( {z_n\over{t_n}}, {y_n\over{t_n}}\right).
\end{equation} 
This means that we are now working in the two-dimensional projective space
{$C\!P_2$}, and that points with homogeneous coordinates $[y,z,t]$ and
$[\lambda \, y, \lambda \, z,\lambda \, t]$ are to be identified
(projectivization).  For (\ref{map}) the map $\varphi$ may be written
as
\begin{equation} \label{projphi}
\varphi: \vek{y}{z}{t} \to \vek{{y}^{3}+a{t}^{3}-{y}^{2}z}
{{y}^{3}\hskip 0.8cm}{t{y}^{2}\hskip 0.8cm}.
\end{equation}

In {$C\!P_2$} the above singularity pattern looks as follows:
\[
\vek0u1 \to \vek100 \to \vek110 \to \vek010 \to \vek000.
\]
The last term of this sequence is not in {$C\!P_2$} and is now the
manifestation of the ambiguity mentioned above. Note also that in this
formulation it is clear that infinities are not singularities: they
look like any other point, the last component is just zero.  The
expansion around the singularity clarifies the situation. We get the
sequence
\begin{eqnarray*}
\vek\epsilon{u}1 &\to& 
\vek{1-u\epsilon^2+\dots}{\epsilon^3\phantom{+\dots}}
	{\epsilon^2\phantom{+\dots}} \to
\vek{1-3u\epsilon^2+\dots}{1-3u\epsilon^2+\dots}{\epsilon^2+\dots} \\
&\to &
\vek{-\epsilon^3+\dots}{1-9u\epsilon^2+\dots}{\epsilon^2+\dots} \\
&\to &
\vek{u\epsilon^8+\dots}{-\epsilon^9+\dots}{\epsilon^8+\dots}
=\vek{u+\dots}{-\epsilon+\dots}{1+\dots},
\end{eqnarray*}
and in the last term we are able to cancel the factor $\epsilon^8$,
after which we can let $\epsilon\to0$, getting $[u,0,1]$. We have thus
emerged from the singularity with the initial information $u$.

The cancellation mentioned here is crucial. It occurs only if there
is a singularity in the map, because the existence of a singularity
means that there will be common factors of $\epsilon$. Such
cancellations of common factors are necessary to reduce the growth of
the degree, because otherwise the successive iterates $\varphi^{(n)}$
of $\varphi$ would be polynomials of degree $d^n$, where $d$ is the
degree of $\varphi$. For integrable systems the cancellations are in
fact so strong that asymptotically the degree grows only polynomially.

The degree of $\varphi$ is not canonical, since it is not invariant
under coordinate changes. However the {\em growth of the degree is
canonical}~\cite{BeVi97}. It is generically exponential~\cite{arnold},
but may become polynomial if the number of common factors is large
enough. The {\em conjecture} is that integrability of the map implies
polynomial growth \cite{FaVi93}(see also~\cite{Ve92}).

For the map (\ref{projphi}) we get the following sequence of degrees:
$ 1,\; 3,\; 9,\; 27,\; 73,\; 195,\; 513,\;1347,\; 3529\dots $. The
first four degrees follow the $3^n$ rule; cancellations then take
place. Note that the first drop of the degree is $3\times 27 - 73 =
8$ corresponding to the factorization of $\epsilon^8$ in the above
calculation.  From the nine first numbers of this sequence we
inferred the generating function for the degrees:
\begin{equation}
g(x) = {\frac {1+3\,{x}^{3}}{\left (1-x\right )\left (1+x\right )\left
({x}^{2}-3\,x+1\right )}}.
\label{g3}
\end{equation}
The next degree, found by iteration of $\varphi$ is 9243 and coincides
with the prediction obtained by expanding $g$:
\begin{eqnarray*}
g & = & 1+3\,x+9\,{x}^{2}+27\,{x}^{3}+73\,{x}^{4}+ 195\,{x}^{5}+
513\,{x}^{6} +  \nonumber\\ 
& & +1347\, {x}^{7}+3529\,{x}^{8}
+9243\,{x}^{9} +24201\,{x}^{10}+\dots
\end{eqnarray*}
(A proof that (\ref{g3}) indeed is the generating function of the
degrees will be given elsewhere~\cite{BeVi97}.)

Function (\ref{g3}) generates a sequence with exponential growth. The
denominator of $g$ contains two basic pieces of information.  It first
shows that the sequence of degrees verifies the very specific relation
\begin{eqnarray}
d_{n+4}- d \cdot d_{n+3} + d \cdot d_{n+1} - d_{n} =0,
\label{recu}
\end{eqnarray}
where $d=d_1=3$ is the degree of the map $\varphi$, and $d_n$ the
actual degree of $\varphi^n$, after ``projectivization''.  It also
determines the asymptotic behavior of $d_n$: if $\alpha$ is the
smallest modulus of the roots of the denominator of (\ref{g3}), then
$d_{n+1} \approx \alpha^{-1} d_n$ asymptotically.  In this case $
\alpha = ({3-\sqrt{5}})/2$ and we define the ``algebraic entropy'' of
the map by
\begin{equation}
{\cal E} \equiv \lim_{n \rightarrow \infty}\frac1n \log(d_n) = \log\left(
{{3+\sqrt{5}}\over{2}}\right).
\end{equation}
This calculation indicates that the map has non-vanishing entropy, and
therefore is likely to be non-integrable.

\par\bigskip
\epsfxsize=\figwidth
\centerline {\epsffile[50 240 562 782]{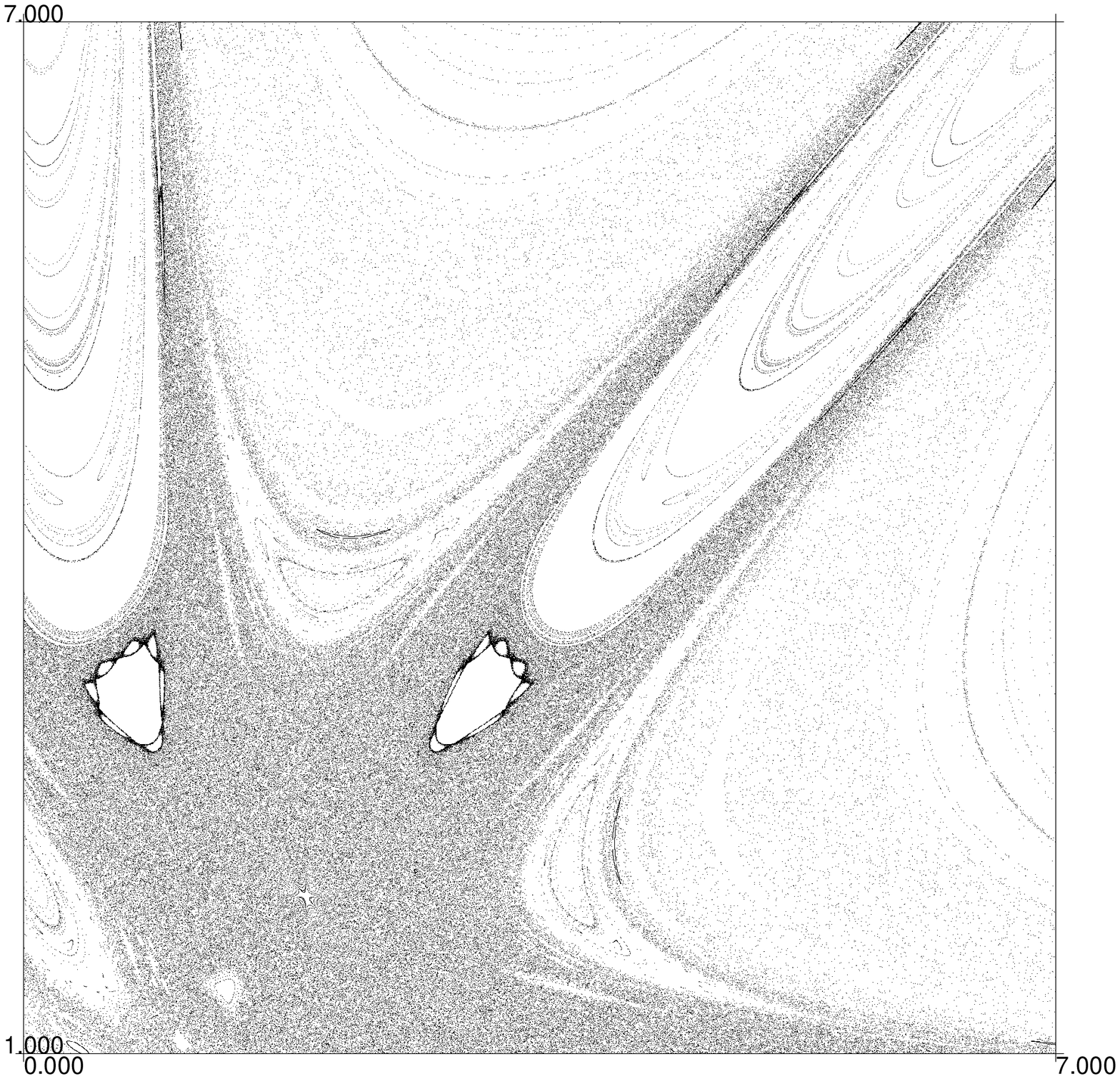} }
\centerline{\it  Fig. 1: A collection of orbits of the map (\ref{map})}
\par \bigskip

Numerical chaos can be seen when we draw a picture of some orbits of
the map, see Figure 1. This figure was obtained with $a=7$.  The two
``cat's paws'' are around two points of a nine-periodic orbit of
$\varphi$. Such a point of order nine of $\varphi$
($x_n=x_{n+1}\simeq 3.043896\dots$) is located at the lower left hand
corner of Figure~2.  The picture is characteristic of chaotic
behavior of a two dimensional conservative system.

In order to compare with a truly integrable system let us consider
\cite{FGR}
\begin{eqnarray}
x_{n+1}+x_{n-1}=\frac{a}{x_n^2}+\frac{b}{x_n},
\end{eqnarray}
which is related to $d-P_I$.  This is also a third degree map, but the
singularity structure is now such that the degrees grow only as $1, 3,
9, 19, 33, 51, 73, 99, 129, 163, 201, \dots$ The generating functional
is
\begin{equation}
g = \frac{1+3x^2}{(1-x)^3},
\label{g2}
\end{equation} 
and in fact the degrees grow polynomially according to the simple rule
\begin{eqnarray*}
d_n = 2n^2+1.
\end{eqnarray*}
 
\par\bigskip
\epsfxsize=\figwidth
\centerline { \epsffile[80 80 592 612]{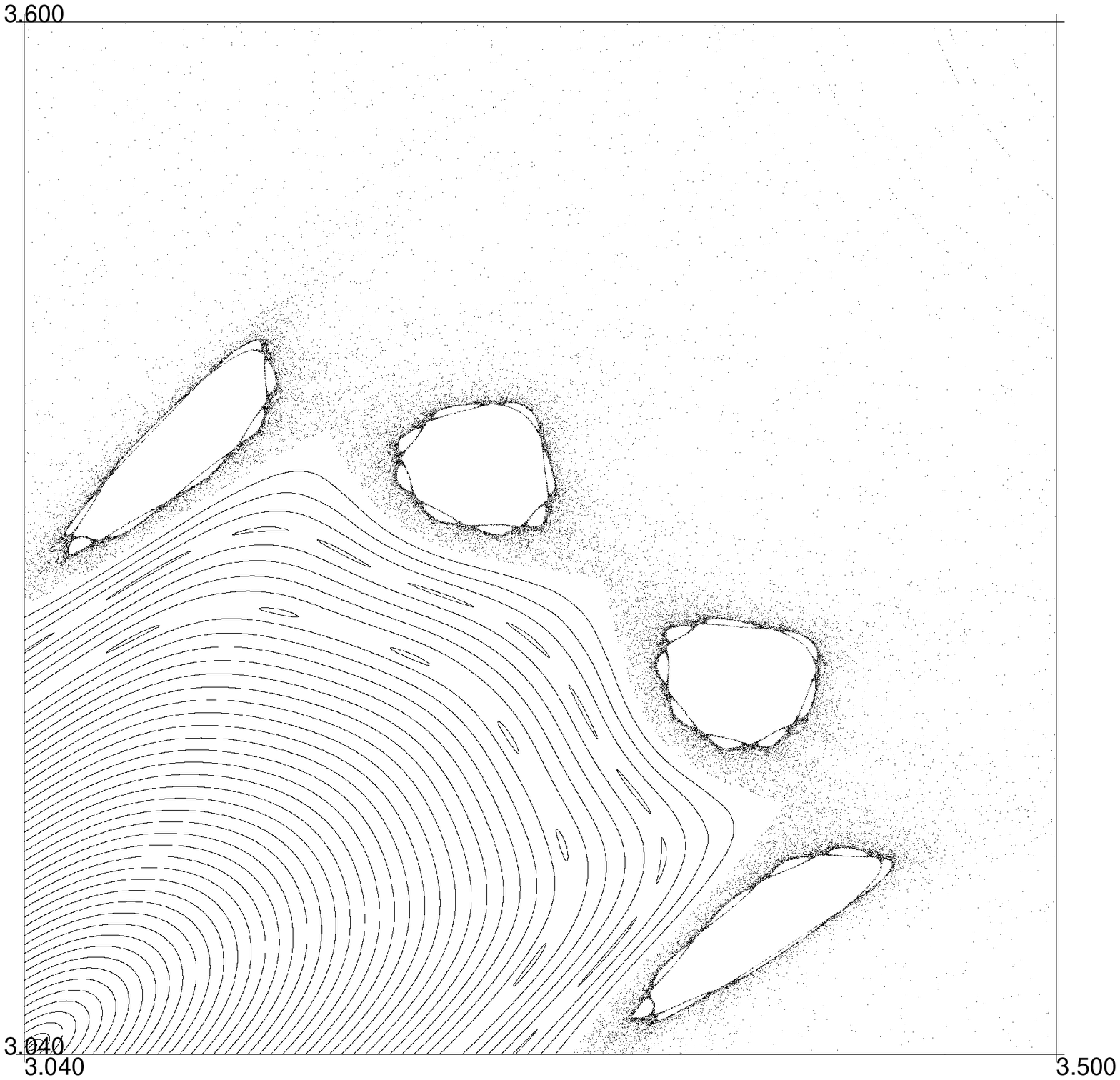} }
\centerline{\it  Fig. 2: A close up of Figure 1}
\par \bigskip

If model (\ref{map}) were an isolated example, it might be dismissed by
some ad hoc rule. However, it turns out that the singularity
confinement test is somewhat insensitive in general, and maps
containing rather arbitrary functions pass the test. Let us consider
the more general map
\begin{equation}
x_{n+1}+x_{n-1}=x_n+f(x_n),
\label{map2}
\end{equation}
where $f$ does not have to be rational. (Indeed, there is no {\it inherent}
reason to limit the singularity confinement test to rational maps,
while the notion of degree, and consequently the definition of the
algebraic entropy is tightly related to (bi)-rationality.  For
non-rational transformations, one should use a definition of entropy
more closely inspired from~\cite{arnold}.)

It is clear that the map (\ref{map2}) can be iterated forwards and
backwards except possibly when $f$ diverges. Let us therefore assume
that $f$ diverges at some points $x_j^*$ with a power series expansion
starting as
\begin{equation}
f(x_j^*+\epsilon)=a_j\,\epsilon^{-K_j}(1+O(\epsilon)),\, K_j> 0,
\end{equation}
and that it vanishes at infinity as
\begin{equation}
f(\epsilon^{-1})=b\,\epsilon^L(1+O(\epsilon)),\, L> 0.
\end{equation}
{}From these assumptions it follows in particular that
$f(f(x_j^*+\epsilon) +O(1)) =ba_j^{-L}\epsilon^{LK_j}(1+O(\epsilon))$.
The singularity analysis proceeds now as follows
\begin{eqnarray*}
& x_{-1} & =u ,\\
& x_0 & =x_j^*+\epsilon,\\
& x_1   &= f(x_j^*+\epsilon)-u+x_j^*+\epsilon,\\
& x_2   &= f(x_j^*+\epsilon)-u+ba_j^{-L}\epsilon^{LK_j}(1+O(\epsilon)),\\
& x_3   &= -x_j^*-\epsilon+2ba_j^{-L}\epsilon^{LK_j}(1+O(\epsilon)),\\
& x_4   &= u-x_j^*+ \Delta_j(\epsilon)  +O(\epsilon),
\end{eqnarray*}
with
\begin{eqnarray*}
\Delta_j(\epsilon) = 
f\left(-x_j^*-\epsilon+2ba_j^{-L}\epsilon^{LK_j}(1+O(\epsilon))\right)-
f(x_j^*+\epsilon).
\end{eqnarray*}
Thus in order for $f$ to pass the singularity confinement test, we only
need to impose the condition that
\begin{equation}
\Delta_j(\epsilon) =O(1),\,\forall j,
\label{E:cc1}
\end{equation}
and that this term does not cancel the $u$ dependence in $x_4$.

A simple calculation shows that (\ref{E:cc1}) is true at least if 1)
$K_j(L-1)\ge 1$ and 2) the singularity structure of $f$ is even, i.e.,
both $+x_j^*$ and $-x_j^*$ are singular points of $f$ and the
expansions at these points match as $f(x_j^*+\epsilon)
-f(-x_j^*-\epsilon) =O(\epsilon)$. The simplest such function is
$x^{-2}$ (yielding the map (\ref{map})), but it is easy to construct
other examples.  If attention is restricted to rational functions, we
can, e.g., pick any two relatively prime polynomials $Q(x)$ and $P(x)$
of degree $M$ and define $f(x)=(P(x)P(-x))/(x^2Q(x)Q(-x))$.  As an
example we performed the degree growth analysis above on the special
case $P=x+5$, $Q=x+3$ and obtained entropy ${\cal E} = \log\left(
(5+\sqrt{21})/2\right)$.  Drawing the orbits again
corroborates the claim of non-integrability.

Another class of maps which passes the singularity confinement test is
contained in
\begin{equation}
x_{n+1}+x_{n-1}=f_n(x_n),
\end{equation}
where we just assume that the functions $f_n$ diverge at some points
$x_j^*$ (independent of $n$), and vanishes at infinity, but it is not
necessary yet to specify how. Here we allow non-autonomous maps, i.e.,
$f$ may also depend on $n$, as indicated by the subscript (of course
this could have been done with the previous model as well). The
singularity analysis now goes as follows:
\begin{eqnarray*}
& x_{-1} & =u,\\
&x_0 &= x_j^*+\epsilon,\\
& x_1 &= -u+f_0(x_j^*+\epsilon),\\
& x_2 &= -x_j^*-\epsilon+
	f_1(f_0(x_j^*+\epsilon)-u),\\
& x_3 & = u-f_0(x_j^*+\epsilon)+f_2\bigl(-x_j^*-
\epsilon+f_1(f_0(x_j^*+\epsilon)-u)\bigr).
\end{eqnarray*}
For the singularity to be confined at this step, it is sufficient that
the behaviors at the singular points and at infinity match so that
\begin{eqnarray}
\forall\,j,n \quad \lim_{\epsilon\to 0} && \left[f_n(x_j^*+\epsilon)\right.
\label{E:cc2}\\ \nonumber - &&\left.
f_{n+2}\bigl(-x_j^*-\epsilon+f_{n+1}(f_{n}(x_j^*+\epsilon)-u)\bigr)\right]=0,
\end{eqnarray}
The singularity pattern is shorter than for (\ref{map}), $u\to 0\to
\infty\to 0 \to \infty-\infty$, and upon expanding the ambiguity
resolves to $u$. The confinement condition (\ref{E:cc2}) is similar to
(\ref{E:cc1}).

The above results show that the singularity confinement test is only
sensitive to the function's behavior at its singular points and at
infinity.  Especially for the non-rational case it is easy to dress a
function which passes the test by something which does not alter this
behavior.

The singularity confinement test is definitely a useful
tool for identifying potentially integrable systems. It is probably
necessary but, in the light of the present results, it appears to be
insufficient. Of course, for a given map, the situation would be
settled if one could establish any of the constructive properties
associated with integrability, such as Lax pair, superposition
principle, conservation laws, but in practice this is very difficult.

It is therefore important to continue developing and refining methods
for algorithmic testing of integrability. For rational maps one such
refinement is presented in this letter: one should look at the growth
of the degree of the map (when written in the projective space): if
the degree grows faster than polynomially (non vanishing ``algebraic
entropy''), it is likely that the system is not integrable.

Part of this work was done when J.H. was visiting the Institute of
Computational Mathematics and Scientific Engineering Computing in
Beijing.  J.H. would like to thank M. Ablowitz, B. Grammaticos,
X.B. Hu, S.H. Wang and Y.X. Yuan for discussions.

\end{document}